\documentclass[twocolumn,showpacs,aps,prd,nobibnotes,nofootinbib,floatfix]{revtex4-2}

\usepackage{amsmath}
\usepackage{graphicx,subfigure}
\usepackage[usenames]{color}
\usepackage[normalem]{ulem}
\usepackage{textcomp}
\usepackage{gensymb}
\usepackage{xspace}
\usepackage{verbatim}

\usepackage{soul}
\usepackage[colorlinks=true]{hyperref}
\hypersetup{
  citecolor  = blue
}
\newcommand{\rh}{r_{\text{h}}}

\begin{document}

\title{First law of black hole thermodynamics and the weak cosmic censorship conjecture for Kerr-Newman Taub-NUT black holes}

\author{Si-Jiang Yang$^a$$^b$}

\author{Wen-Di Guo$^a$$^b$}

\author{Shao-Wen Wei$^a$$^b$}

\author{Yu-Xiao Liu$^a$$^b$$^c$}%
\email{liuyx@lzu.edu.cn, corresponding author}
\affiliation{$^{a}$Lanzhou Center for Theoretical Physics, Key Laboratory of Theoretical Physics of Gansu Province, and Key Laboratory of Quantum Theory and Applications of MoE, Lanzhou University, Lanzhou, Gansu 730000, China\\
$^{b}$Institute of Theoretical Physics $\&$ Research Center of Gravitation, Lanzhou University, Lanzhou 730000, China}
\date{\today}

\begin{abstract}
Stimulated by the recent researches of black hole thermodynamics for black hole with Newman-Unti-Tamburino (NUT) parameter, we investigate the thermodynamics and weak cosmic censorship conjecture for a Kerr-Newman Taub-NUT black hole. By defining the electric charge as a Komar integral over the event horizon, we construct a consistent first law of black hole thermodynamics for a Kerr-Newman Taub-NUT black hole through Euclidean action. Having the first law of black hole thermodynamics, we investigate the weak cosmic censorship conjecture for the black hole with a charged test particle and a complex scalar field. We find that an extremal black hole cannot be destroyed by a charged test particle and a complex scalar field. For a near-extremal black hole with small NUT parameter, it can be destroyed by a charged test particle but cannot be destroyed by a complex scalar field. 
\end{abstract}
\maketitle

\section{Introduction}\label{sec:intro}

Gravitational collapse inevitably leads to spacetime singularity. This is the famous Haking-Penrose singularity theorem~\cite{Penr65,HaPe70}. The presence of spacetime singularity might indicate the failure of general relativity in the area of  extremely strong Gravitational field. To preserve the predictability of gravitational theory, Penrose proposed the weak cosmic censorship conjecture which states that spacetime singularities are hidden behind black hole event horizon and can never be seen by distant observers~\cite{Penr69}. The weak cosmic censorship conjecture preserves the predictability of gravitational theory outside the event horizon, but it also forbids us to probe physics near the singularity where quantum gravity effects cannot be neglected. The weak cosmic censorship conjecture has been proposed more than 50 years, but a general proof of the conjecture is still out of reach.

Though we cannot give a general proof of the conjecture, we can still check it. There are many ways to test the weak cosmic censorship conjecture. The one we are interested in here is to test the conjecture through gedanken experiment.

The idea of testing the conjecture through thought experiment was first proposed by Wald~\cite{Wald74}, who proposed to test the weak cosmic censorship conjecture by throwing test particles with large charge or angular momentum into an extremal Kerr-Newman black hole, and found that particles causing the destruction of  the event horizon cannot be captured by the extremal black hole due to the Coulomb and centrifugal repulsion force. Systematic works of Rocha and Cardoso et al. for Ba\~nados-Teitelboim-Zanelli (BTZ) black hole~\cite{RoCa11}, higher-dimensional Myers-Perry family of rotating black holes and a large class of five-dimensional black rings~\cite{BCNR10} also suggested that extremal black holes cannot be destroyed by test particle. But further investigations suggest that a near-extremal black hole can be destroyed by test particles~\cite{Hod02,Hube99,JaSo09}. Furthermore, the work of Gao and Zhang showed that even an extremal Kerr-Newman black hole can be destroyed if we take into account the second order of the energy, angular momentum, and charge of the charged test particle~\cite{GaZh13}. Similar counterexamples are found in modified theories of gravity~\cite{GhFS19,GhMS21}. But when backreaction and self-force are taken into account, these counterexamples seemed to be rescued~\cite{BaCK10,BaCK11,ZVPH13,CoBa15,Gwak17,LiWL19}. Besides testing the conjecture by throwing test particle, we can also check it through the scattering of fields~\cite{Semi11,Gwak18,Gwak21,YWCYW20,LGCM21,Gwak20,YZWL22}. Usually, classical fields cannot destroy the event horizon. Recently, Sorce and Wald proposed a new thought experiment by taking into account the second order perturbations from the matter fields, and they found that a near-extremal Kerr-Newman black hole cannot be destroyed~\cite{SoWa17}. Subsequent systematic works further support the result that the event horizon cannot be destroyed by this kind of gedanken experiment~\cite{SaJi21,ZhJi20,QYWR20,WaJi20,JiGa20,ChLN19,QuTW22,CLNC21,ShNa22,HTWZ22}.

A black hole with NUT parameter arises puzzling questions. It carries a peculiar gravitational charge, namely the NUT charge, which is very similar to the magnetic monopole~\cite{HeKM19}. Though solutions of Einstein field equation with NUT parameter were obtained in the early 1950s~\cite{Taub51,NeTU63}, they were even not regarded as black holes due to the strange properties of the Misner strings~\cite{Misn63}. Recent researches suggest that solutions with NUT parameter are not so strange as previous thought. Contrary to previous doubts~\cite{Holz06,KeMn06}, various viewpoints on consistent thermodynamics for black hole with NUT parameter have been formulated~\cite{HeKM19,BoGK19,BGHK19,Durk22,ClGa20,WuWu19,ChJi19,AbTM21,Frod22,LiLM22}.

By introducing a pair of conjugate thermodynamic variables, the NUT charge $N$ and the Misner potential $\psi$, Hennigar et al. proposed a consistent first law for Lorentz Taub-NUT black hole~\cite{HeKM19}. Following this way, Ballon Bordo et al. obtained two kinds of consistent first laws of thermodynamics for rotating NUTty dyons through different definitions for electric charge and magnetic charge~\cite{BaGK20}. By choosing different gauge for the electric and magnetic potentials, Ballon Bordo et al. got the electric and magnetic first laws of black hole thermodynamics, which correspond to $g=-er/\left(\rh^2+a^2-n^2  \right)$ and $e=4gn^2\rh/\left(\rh^2+a^2-n^2  \right)$, respectively. Here $e$ and $g$ are the electric and magnetic parameters, respectively. However, they did not provide the first law of black hole thermodynamics for the case of vanishing magnetic parameter. In this case, the rotating NUTty dyons reduce to the Kerr-Newman Taub-NUT black hole. There are also multi-hair viewpoint~\cite{WuWu19} for the thermodynamics for Kerr-Newman Taub-NUT black hole. But it leads mathematically problem that the mass and the NUT parameter should be interpreted as three independent thermodynamic variables.  Our first task is to get a consistent thermodynamics for the Kerr-Newman Taub-NUT black hole.

Many works suggest that there are close relationship between black hole thermodynamics and weak cosmic censorship conjecture~\cite{LiNC22,NaQV16,GoNa20,AEKM23}. As our previous work suggested~\cite{YCWWL20} that if we do not take the thermodynamics into account or using inappropriate thermodynamics, this will lead to the result that both extremal and near-extremal black holes can be destroyed by a scalar field~\cite{Duzt18}. Using our obtained first law of black hole thermodynamics, we investigate the weak cosmic censorship conjecture for Kerr-Newman Taub-NUT black holes. We find that an extremal Kerr-Newman Taub-NUT cannot be destroyed by a charged test particle. While a near-extremal black hole with small NUT parameter can be destroyed by a charged test particle. However, for charged scalar field scattering, both extremal and near-extremal black holes cannot be destroyed.

The outline of the paper is as follows. In Sec.~\ref{2}, we investigate the first law of black hole thermodynamics for Kerr-Newman Taub-NUT black hole. In Sec.~\ref{3} and Sec.~\ref{4}, we try to destroy the event horizon of the extremal and near-extremal Kerr-Newman Taub-NUT black holes by a charged test particle and a charged scalar field, respectively. The last section is devoted to discussion and conclusion.

\section{Kerr-Newman Taub-NUT black hole and its thermodynamics}\label{2}

The Kerr-Newman Taub-NUT spacetime is a four-dimensional electrovacuum solution of the Einstein's field equation~\cite{PlDe76}.
The metric for the Kerr-Newman Taub-NUT spacetime in Boyer-Lindquist coordinates can be written in the form
\begin{equation}
  \begin{split}
     ds^2 &=-\frac{\Delta}{U}\left[\mathrm{d}t+\left(2n\cos\theta-a\sin^2\theta\right)\mathrm{d}\phi\right]^2 +
  \frac{U}{\Delta}\mathrm{d}r^2  \\
  &+U\mathrm{d}\theta^2
  +\frac{\sin^2\theta}{U}\left[a\mathrm{d}t
  -\left(r^2+n^2+a^2\right)\mathrm{d}\phi\right]^2,
  \end{split}
\end{equation}
and the electromagnetic field potential
\begin{equation}\label{Apotential}
  \mathrm{A}=-\frac{er}{U}\mathrm{d}t-\frac{(2n\cos\theta-a\sin^2\theta)er}{U}\mathrm{d}\phi,
\end{equation}
which leads to the electromagnetic field
\begin{equation}\label{Etensor}
  \begin{split}
     F & =\mathrm{d}\mathrm{A} =\frac{\left[r^2-(n+a \cos\theta)^2\right]e}{U^2}\mathrm{d}r\wedge \mathrm{d}t \\ &-\frac{2\left(n+a\cos\theta\right)ear\sin\theta}{U^2}\mathrm{d}\theta\wedge \mathrm{d}t+ \\
       &  \frac{\left[r^2-\left(n+a\cos\theta\right)^2\right] \left(2n\cos\theta-a\sin^2\theta\right)e}{U^2}\mathrm{d}r\wedge \mathrm{d}\phi \\
       &+\frac{2\left(r^2+n^2+a^2\right)\left(n+a\cos\theta\right)er\sin\theta}{U^2}\mathrm{d}\theta\wedge \mathrm{d}\phi.
  \end{split}
\end{equation}
Here the metric functions are given by
\begin{align}\label{Mfunction}
  \Delta & =r^2-2mr+a^2+e^2-n^2, \\
  U & =r^2+(n+a\cos\theta)^2,
\end{align}
where $m$, $a$, $e$ and $n$ are the mass parameter, the angular momentum parameter, the electric parameter, and the NUT parameter, respectively.

\begin{figure}
  \centering
  \includegraphics[width=9cm]{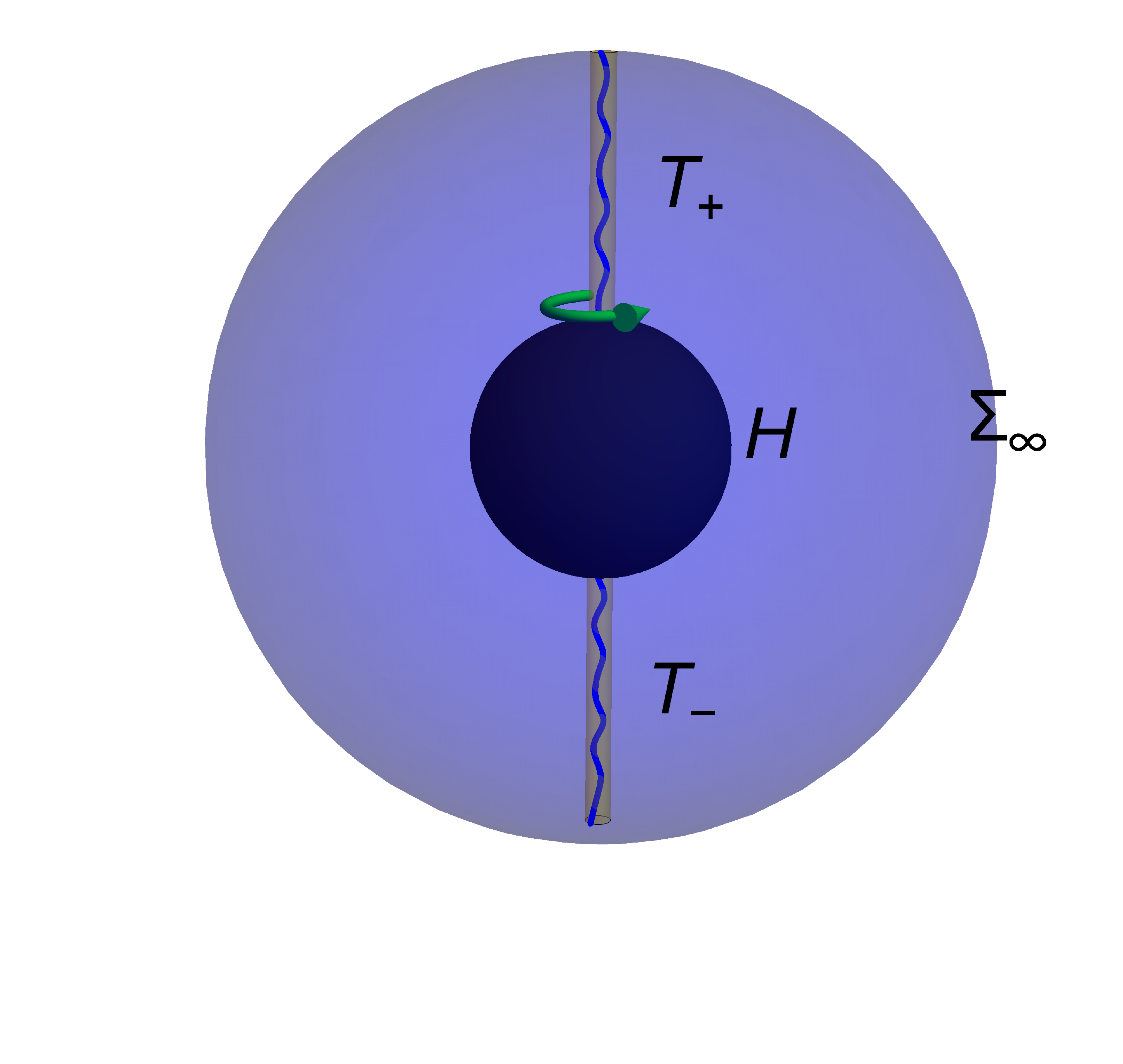}
  \caption{Boundaries of the Kerr-Newman Taub-NUT black hole~\cite{BGHK19a,YCWWL20}: Misner tubes. Apart from the standard boundaries, the event horizon $H$ and spatial infinity $\Sigma_\infty $, the Kerr-Newman Taub-NUT spacetime has two Misner tubes $T_{\pm}$ located at the north and south poles.}\label{KNTBHfig}
\end{figure}

The above metric describes a rotating black hole with two Misner strings symmetrically located on the north and south poles as depicted in Fig.~\ref{KNTBHfig}.
Due to the presence of the Misner strings, the spacetime is not asymptotical flat. For vanishing NUT parameter, the metric describes a Kerr-Newman black hole. For nonvanishing NUT parameter, there are two string singularities corresponding to the two Misner strings. 
Besides the string singularities and horizon, there is a spacetime singularity for $ n\leq a $. For $ n\leq a $, there is a ring singularity located at $r=0$ and $ \cos\theta=n/a$. However, the spacetime is regular and there is no spacetime singularity for $n>a$, and the metric describes a nonsingular black hole.

The horizons of the black hole are determined by
\begin{equation}\label{HorizonE}
  \Delta =r^2-2mr+a^2+e^2-n^2=0.
\end{equation}
Solving the quadratic equation for $r$, we obtain the horizons of the black hole
\begin{equation}\label{EventH}
  r_{\pm}=m\pm \sqrt{m^2+n^2-a^2-e^2},
\end{equation}
with the plus sign corresponds to the event horizon, and the minus sign corresponds to the inner horizon, which is also the Cauchy horizon. In the following, we use $\rh$ to denote the event horizon.

The spacetime is stationary and axisymmetric. The mass of the black hole is
\begin{equation}\label{Mass}
  M=m=\frac{\rh^2+a^2+e^2-n^2}{2\rh}.
\end{equation}
The area of the black hole event horizon is
\begin{equation}\label{Area}
  A=4\pi (\rh^2+a^2+n^2).
\end{equation}
The Bekenstein-Hawking entropy is different from the Noether entropy for black hole with NUT parameter~\cite{HeKM19}, and there is no consensus on which definition for entropy should be used. Here, we use the Benkenstein-Hawking entropy as the black hole  entropy. The entropy for the black hole is
\begin{equation}\label{Entropy}
  S=\frac{A}{4}=\pi (\rh^2+a^2+n^2).
\end{equation}

The metric describes a rotating charged black hole with angular velocity
\begin{equation}\label{angularV}
  \Omega_{\text{h}}=\frac{a}{\rh^2+a^2+n^2},
\end{equation}
and electrostatic potential
\begin{equation}\label{Epotetional}
  \phi_{\text{h}}=\frac{e\rh}{\rh^2+a^2+n^2}.
\end{equation}
The electric charge surrounded by the event horizon is
\begin{equation}\label{EchargeH}
\begin{split}
    Q  &= \frac{1}{4\pi}\int_{\text{H}}*F \\
     & = \frac{e \left(2 a^2 \rh^2+a^4-n^4+\rh^4\right)}{\left(a^2-2 a n+n^2+\rh^2\right) \left(a^2+2 a n+n^2+\rh^2\right)}.
\end{split}
\end{equation}

The event horizon of the black hole is the same as the Killing horizon with Killing vector
\begin{equation}\label{KillingH}
  K=\partial_t+\Omega_{\text{h}}\partial_\phi.
\end{equation}
The surface gravity of the black hole event horizon is
\begin{equation}\label{SufaceG}
  \kappa=\frac{\Delta'(\rh)}{2(\rh^2+n^2+a^2)}=\frac{1}{2 \rh}\frac{\rh^2+n^2-a^2-e^2}{\rh^2+n^2+a^2}.
\end{equation}
Then, the temperature associated with the black hole event horizon reads
\begin{equation}\label{Tempe}
  T=\frac{\kappa}{2\pi}=\frac{1}{4\pi \rh}\frac{\rh^2+n^2-a^2-e^2}{\rh^2+n^2+a^2}.
\end{equation}

Besides the black hole horizon, there are also two Misner string horizons, which are Killing horizons correspond to the Killing vector fields
\begin{equation}\label{MisnerH}
  \xi_{\pm}=\partial_t\mp \frac{1}{2n}\partial_{\phi}.
\end{equation}
The surface gravity for the Misner string horizons can be calculated by the standard formula
\begin{equation}\label{SurfaceF}
  \kappa_{\pm}^2=\frac{1}{2}\nabla_\mu \xi_\nu \nabla^\mu \xi^\nu,
\end{equation}
which gives the result
\begin{equation}\label{SurfaceGMh}
  \kappa_{\pm}=\frac{1}{2n}.
\end{equation}
As in the work of Hennigar et al.~\cite{HeKM19}, the surface gravity of the two Misner string horizons is interpreted as the Misner potential
\begin{equation}\label{Mpotential}
  \psi=\frac{\kappa_{\pm}}{2\pi}=\frac{1}{8\pi n}.
\end{equation}

To get the first law of black hole thermodynamics for the Kerr-Newman Taub-NUT black hole, we consider the action  for the Kerr-Newman Taub-NUT AdS spacetime by using the standard AdS counterterms, and take the asymptotically flat limit. The Euclidean action for the Kerr-Newman Taub-NUT AdS spacetime is~\cite{BaGK20}
\begin{multline}\label{EuclideanAct}
  I=\frac{1}{16\pi}\int_Md^4x\sqrt{g}\left(R+\frac{6}{l^2}-F^2\right)\\
  +\frac{1}{8\pi}\int_{\partial M}d^3x\sqrt{h}\left(\mathcal{K}-\frac{2}{l}-\frac{l}{2}\mathcal{R}\right),
\end{multline}
where $\mathcal{K}$ and $\mathcal{R}$ are the extrinsic curvature and Ricci scalar of the boundary respectively, and $h$ is the determinant of the induced metric. The action is related to the corresponding free energy $G=I/\beta$ with $\beta=1/T$. After taking the asymptotically flat limit $l\to +\infty$, the free energy for the Kerr-Newman Taub-NUT black hole is
\begin{equation}\label{GibbFE}
  G=m-\frac{\rh}{2}\frac{(\rh^2-n^2+a^2)e^2}{[\rh^2+(a+n)^2][\rh^2+(a-n)^2]}.
\end{equation}

The Gibbs free energy $G$ can be regarded as a function of the temperature $T$, the Minsner potential $\psi$, the angular velocity $\Omega_{\text{h}}$ and the electric potential $\phi_h$, with:
\begin{equation}\label{Gfunction}
  G=G(T,\Omega_{\text{h}},\phi_{\text{h}},\psi),
\end{equation}
with which we can define the following thermodynamic quantities:
\begin{align}\label{DThermodquant}
  S &=-\frac{\partial G}{\partial T}, ~~Q=-\frac{\partial G}{\partial \phi_{\text{h}}}, ~~J= -\frac{\partial G}{\partial \Omega_{\text{h}}}, ~~N=-\frac{\partial G}{\partial \psi}.
\end{align}
From the above definition, we can get the thermodynamical quantities:
\begin{equation}\label{Entr}
  S = \pi (\rh^2+a^2+n^2),
\end{equation}
\begin{equation}\label{Echarge}
  Q  = \frac{e \left(2 a^2 \rh^2+a^4-n^4+\rh^4\right)}{\left(a^2-2 a n+n^2+\rh^2\right) \left(a^2+2 a n+n^2+\rh^2\right)},
\end{equation}
\begin{equation}\label{Angulam}
  \begin{split}
     J & =  \frac{1}{2} \left[\frac{a \left(a^2+n^2\right) \left(a^2+e^2-n^2\right)}{\rh (a-n) (a+n)}+\frac{2 e^2 n^2 \rh (a-n)}{\left((a-n)^2+\rh^2\right)^2} \right.\\
     &+\frac{2 e^2 n^2 \rh (a+n)}{\left((a+n)^2+\rh^2\right)^2} -\frac{e^2 n^2 \rh}{\rh^2 (a-n)+(a-n)^3}\\
       & \left. -\frac{e^2 n^2 \rh}{\rh^2 (a+n)+(a+n)^3}+a \rh\right],
  \end{split}
\end{equation}
\begin{equation}\label{Ncharge}
  \begin{split}
     N & =4\pi n^3 \left[\frac{( n^2-a^2 ) (4 a^4 + 7 a^2 e^2 +
    5 e^2 n^2 - 4 n^4) \rh}{ \left(a^4 + 2 a^2 (\rh^2-n^2 ) + (n^2 + \rh^2)^2\right)^2} \right.  \\
       & +\frac{(a^2 + e^2 - n^2) (n^2-a^2 )^3 -\rh^6 \left(4 a^2-3 e^2+4 n^2\right)}{\rh (a^4 + 2 a^2 (\rh^2-n^2 ) + (n^2 + \rh^2)^2)^2} \\
       & \left.-\frac{\rh^7+\rh^3 \left(a^2 \left(3 e^2+4 n^2\right)+6 a^4-7 e^2 n^2+6 n^4\right)}{ \left(a^4+2 a^2 \left(\rh^2-n^2\right)+\left(n^2+\rh^2\right)^2\right)^2}\right],
  \end{split}
\end{equation}
where the electric charge is defined as the charge surrounded by the event horizon instead of infinity.

It is evident that the above thermodynamic quantities satisfy the first law of black hole thermodynamics and the Smarr relation:
\begin{align}
  dM= & TdS+\Omega_{\text{h}}dJ+\phi_{\text{h}}dQ+\psi dN, \\
  M =& 2\left(TS+\Omega_{\text{h}}J+\psi N\right)+\phi_{\text{h}}Q.
\end{align}

From the first law of black hole thermodynamics for the Kerr-Newman Taub-NUT black hole, if the black hole absorbs a particle or field with energy $\delta E$, angular momentum $\delta J$ and charge $\delta Q$, we can get the change of the black hole parameters from the first law of black hole thermodynamics. Evidently, the NUT charge is conserved during the absorption process as our previous works indicated~\cite{YCWWL20,FYTYL21}. Then we can check the validity of the weak cosmic censorship conjecture for the Kerr-Newman Taub-NUT black hole.

\section{Destroying the black hole with charged test particles}\label{3}

The viewpoint of gedanken experiment to destroy the event horizon of a black hole was first proposed by Wald in 1974~\cite{Wald74}. By dropping a test particle with large charge or angular momentum into an extremal Kerr-Newman black hole, Wald found that particles causing the destruction of the event horizon can not be absorbed by the black hole~\cite{Wald74}. But further investigations show that a near-extremal black hole can be destroyed by a test particle with large charge or angular momentum~\cite{Hube99,Hod02,JaSo09}.

To check the validity of the weak cosmic censorship conjecture for the Kerr-Newman Taub-NUT black hole, we shoot a test particle with large charge or/and angular momentum into an extremal or near-extremal black hole. The equation of motion for a particle with rest mass $M_0$ and charge $\delta Q$ in the Kerr-Newman Taub-NUT spacetime is
\begin{equation}\label{EqMotion}
  \frac{d^2x^\mu}{d\tau^2}+\Gamma^\mu_{\alpha\beta}\frac{dx^\alpha}{d\tau}\frac{dx^\beta}{d\tau}=\frac{\delta Q}{M_0}F^\mu_
  {~~\nu} \frac{dx^\nu}{d\tau}.
\end{equation}

The equation of motion for the charged test particle in the Kerr-Newman Taub-NUT spacetime can also be derived from the Lagrangian
\begin{equation}\label{Lagrangian}
\begin{split}
  L&=\frac{1}{2}M_0g_{\mu\nu}\frac{dx^\mu}{d\tau}\frac{dx^\nu}{d\tau}+\delta Q A_\mu \frac{dx^\mu}{d\tau}  \\
  &=-\frac{1}{2}M_0g_{\mu\nu}\dot{x}^\mu\dot{x}^\nu+\delta Q A_\mu \dot{x}^\mu.
  \end{split}
\end{equation}

We shoot a charged test particle into the equatorial plane of the black hole, with an angular momentum $\delta J$, aligned in the same direction as the black hole. From the Lagrangian for the test charged particle, we can get the energy $\delta E$ and angular momentum $\delta J$, which are
\begin{subequations}\label{EandJ}
  \begin{align}
    \delta E & =-P_t=-\frac{\partial L}{\partial \dot{t}}=-M_0g_{0\nu}\dot{x}^\nu-\delta Q A_t,  \label{EnergyPa} \\
     \delta J &= P_\phi=\frac{\partial L}{\partial \dot{\phi}}=M_0g_{3\nu}\dot{x}^\nu+\delta Q A_\phi, \label{AngularmonmentumParticle} \\
          P_\theta & =\frac{\partial L}{\partial \dot{\theta}}=0.
  \end{align}
\end{subequations}
For a stationary spacetime, the energy $\delta E$ is a constant of motion; for an axisymmetric spacetime, the angular momentum $\delta J$ is a constant of motion. Since the Kerr-Newman Taub-NUT spacetime is stationary and axisymmetric, both the energy $\delta E$ and angular momentum $\delta J$ of the particle are constants of motion in the spacetime.

From the four-velocity of the particle, we obtain
\begin{equation}\label{FourV}
  U^\mu U_\mu=\frac{1}{M_0^2}g^{\alpha\beta}\left(P_\alpha-A_\alpha \delta Q\right)\left(P_\beta-A_\beta\delta Q\right)=-1.
\end{equation}
Therefore, the energy $\delta E$ and angular momentum $\delta J$ of the particle satisfy the following equation:
\begin{multline}\label{DEandDJ}
g^{00}\left(\delta E+A_t \delta Q\right)^2 -2g^{03}\left(P_\phi-A_\phi \delta Q\right)\left(\delta E+A_t \delta Q\right)\\
  +g^{11}P_r^2+g^{33}\left(P_\phi-A_\phi \delta Q\right)^2+M_0^2=0.
\end{multline}
The above equation is a quadratic equation of energy $\delta E$. Solving the quadratic equation, we can get the energy of the particle,
\begin{multline}\label{Energy}
 \delta E  =-A_t \delta Q+\frac{g^{03}}{g^{00}}\left(P_\phi-A_\phi \delta Q\right)\pm\frac{1}{g^{00}}\left\{  \left(g^{03} \right)^2 \left(P_\phi-\right.\right. \\
  \left.\left. A_\phi \delta Q\right)^2 -g^{00}\left[ g^{11} P_r^2 +g^{33}\left(P_\phi-A_\phi \delta Q\right)^2+M_0^2\right]\right\}^{\frac{1}{2}}.
\end{multline}
The motion of the particle should be future-directed. It follows that
\begin{equation}\label{FutureDpm}
  \frac{dt}{d\tau}>0.
\end{equation}
The future-directed condition for the particle implies that the energy of the particle ought to be
\begin{multline}\label{Energy}
 \delta E  =-A_t \delta Q+\frac{g^{03}}{g^{00}}\left(P_\phi-A_\phi \delta Q\right) -\frac{1}{g^{00}}\left\{  \left(g^{03} \right)^2 \left(P_\phi- \right.\right. \\
    \left.\left. A_\phi \delta Q\right)^2 -g^{00}\left[ g^{11} P_r^2 +g^{33}\left(P_\phi-A_\phi \delta Q\right)^2+M_0^2\right]\right\}^{\frac{1}{2}}.
\end{multline}
From Eqs.~\eqref{EnergyPa} and~\eqref{AngularmonmentumParticle}, we have
\begin{subequations}
    \begin{align}
    M_0g_{00}\dot{t}+M_0g_{03}\dot{\phi}&=-\delta E-\delta Q A_t,  \label{EnergyP} \\
     M_0g_{30}\dot{t}+M_0g_{33}\dot{\phi} &=\delta J -\delta Q A_\phi. \label{AngularM}
  \end{align}
\end{subequations}
Clearly, the above two equations are linear equations for $\dot{t}$ and $ \dot{\phi}$. By utilizing Eq.~\eqref{EnergyP} and Eq.~\eqref{AngularM}, we can deduce the following result:
\begin{equation}\label{dott}
  \dot{t}\equiv \frac{dt}{d\tau}=-\frac{\left(\delta E+A_t \delta Q\right)g_{33}+\left(\delta J-A_\phi \delta Q\right)g_{03}}{M_0\left(g_{00}g_{33}-g_{03}^2\right)}.
\end{equation}
Imposing the future-directed condition on the motion of the particle, we obtain
\begin{equation}\label{condition1}
\left(\delta E+A_t \delta Q\right)g_{33}+\left(\delta J-A_\phi \delta Q\right)g_{03}>0.
\end{equation}
Thus, we have
\begin{equation}\label{Condition1b}
  \delta E>\left(\frac{g_{03}}{g_{33}}A_\phi-A_t\right)\delta Q-\frac{g_{03}}{g_{33}}\delta J.
\end{equation}
If the particle can be captured by the black hole, it will inevitably cross the event horizon at a certain point along its trajectory. At the precise moment of crossing the event horizon, the aforementioned condition takes the form of
\begin{equation}\label{AbsorpCond}
\begin{split}
   \delta E &  >\frac{e\rh}{\rh^2+n^2+a^2}\delta Q+\frac{a}{\rh^2+n^2+a^2}\delta J \\
     &=\phi_{\text{h}}\delta Q+ \Omega_{\text{h}}\delta J.
\end{split}
\end{equation}
Evidently, if the angular momentum or/and charge of the test particle is too large, the particle just ``miss'' the black hole and cannot be absorbed by the black hole due to the Coulomb and centrifugal repulsion force.

After the absorption of the charged test particle, the parameters of the black hole changes as
\begin{equation}\label{ParameterChange}
  \begin{split}
     M &\rightarrow M'=M+\delta E,  ~~~~~~~~~~~J \rightarrow J'=J+\delta J,     \\
      Q & \rightarrow Q'=Q+\delta Q, ~~~~~~~~~~~N \rightarrow N'=N.
  \end{split}
\end{equation}
and the minimal of the metric function of the black hole changes as
\begin{equation}\label{MFunc}
  \Delta_{\text{min}}(M,J,Q,N) \rightarrow \Delta'_{\text{min}}(M',J',Q',N').
\end{equation}
After the absorption of the charged test particle, the minimal of the metric function is
\begin{equation}\label{MFunction}
\begin{split}
  \Delta'_{\text{min}}&=\Delta_{\text{min}}(M+\delta E, J+\delta J, Q+\delta Q, N) \\
  &=\Delta_{\text{min}}+\frac{\partial \Delta_{\text{min}}}{\partial M}\delta E+\frac{\partial \Delta_{\text{min}}}{\partial J}\delta J+\frac{\partial \Delta_{\text{min}}}{\partial Q}\delta Q.
  \end{split}
\end{equation}
To destroy the event horizon of the black hole, we only need the minimal of the metric function to be positive,
\begin{equation}\label{DCond}
  \begin{split}
  \Delta'_{\text{min}}&=\Delta_{\text{min}}+\frac{\partial \Delta_{\text{min}}}{\partial M}\delta E+\frac{\partial \Delta_{\text{min}}}{\partial J}\delta J+\frac{\partial \Delta_{\text{min}}}{\partial Q}\delta Q \\
  &=\frac{\partial \Delta_{\text{min}}}{\partial M}\left(\delta E+\frac{   \frac{\partial \Delta_{\text{min}}}{\partial J}   }{    \frac{\partial \Delta_{\text{min}}}{\partial M}  }\delta J
  +\frac{   \frac{\partial \Delta_{\text{min}}}{\partial Q}  }{   \frac{\partial \Delta_{\text{min}}}{\partial M} }\delta Q
  +\frac{\Delta_{\text{min}} }{   \frac{\partial \Delta_{\text{min}}}{\partial M}   }\right)\\
  & >0.
  \end{split}
\end{equation}
Consequently, we can get an upper bound for the energy of the charged test particle, which is
\begin{equation}\label{UpBound}
\begin{split}
  \delta E& <-\frac{    \frac{\partial \Delta_{\text{min}}}{\partial J}    }{  \frac{\partial \Delta_{\text{min}}}{\partial M}    }\delta J
  -\frac{ \frac{\partial \Delta_{\text{min}}}{\partial Q}    }{   \frac{\partial \Delta_{\text{min}}}{\partial M}     }\delta Q
  -\frac{\Delta_{\text{min}} }{    \frac{\partial \Delta_{\text{min}}}{\partial M}     } \\
     & \equiv \Omega_{\text{eff}}\delta J+\Phi_{\text{eff}}\delta Q-E_0.
\end{split}
\end{equation}
where we have defined the effective angular velocity $\Omega_{\text{eff}}$, effective electric potential $\Phi_{\text{eff}}$ and energy $E_0$ as
\begin{equation}\label{effangularVPE}
  \begin{split}
     \Omega_{\text{eff}}&=-\frac{\left(    \frac{\partial \Delta_{\text{min}}}{\partial J}   \right)_{M,Q,N}  }{ \left(    \frac{\partial \Delta_{\text{min}}}{\partial M}     \right)_{J,Q,N}   }, ~~~~~
     \Phi_{\text{eff}}= -\frac{\left(    \frac{\partial \Delta_{\text{min}}}{\partial Q}   \right)_{M,J,N}  }{ \left(
      \frac{\partial \Delta_{\text{min}}}{\partial M}     \right)_{J,Q,N}   }, \\
     E_0&=\frac{\Delta_{\text{min}} }{ \left(    \frac{\partial \Delta_{\text{min}}}{\partial M}     \right)_{J,Q,N}   }.
  \end{split}
\end{equation}
Only when the conditions~\eqref{AbsorpCond} and~\eqref{UpBound} are satisfied simultaneously, can the black hole be destroyed by the charged test particle.

For an extremal Kerr-Newman Taub-NUT black hole, Eq.~\eqref{DCond} becomes
\begin{equation}\label{ExteDestroyp}
\begin{split}
   \Delta'_{\text{min}}&= -\frac{\left( m^2+n^2+a^2 \right)\Theta}{\Upsilon}\left[\delta E-\frac{a}{m^2+n^2+a^2}\delta J \right. \\
   &\left.-\frac{m\sqrt{m^2+n^2-a^2}}{m^2+n^2+a^2}\delta Q    \right]>0,
\end{split}
\end{equation}
where $\Upsilon$ and $\Theta$ are defined as
\begin{equation}\label{DUpsi}
  \begin{split}
     \Upsilon & = 3 m^2 (a^2 + m^2)^2 (a^4 - 6 a^2 m^2 + m^4) \\
       & +
  2 (a^2 + m^2)^2 (2 a^4 - 19 a^2 m^2 + 13 m^4) n^2  \\
       & +
  4 (11 m^6-3 a^6 - 3 a^4 m^2 + 17 a^2 m^4 ) n^4 \\
       &  +
  6 (2 a^4 + 9 a^2 m^2 + m^4) n^6 - (4 a^2 + 15 m^2) n^8,
  \end{split}
\end{equation}
\begin{equation}\label{DTheta}
  \begin{split}
     \Theta & =2 m \left[8 m^2 n^4 \left(9 a^2+2 m^2\right) -3n^8 \right.\\
     &+ 3 \left(a^2+m^2\right)^2 \left(a^4-6 a^2 m^2+m^4\right) \\
       &\left. -2 n^2 (a^2-m^2) \left(30 a^2 m^2+3 a^4+11 m^4-3n^4 \right)  \right].
  \end{split}
\end{equation}

Evidently, the condition~\eqref{ExteDestroyp} to destroy an extremal Kerr-Newman Taub-NUT black hole can be simplified as
\begin{equation}\label{ExteDestroySim}
  \begin{split}
     \delta E & < \frac{a}{m^2+a^2+n^2}\delta J+\frac{me}{m^2+a^2+n^2}\delta Q \\
       & =\Omega_{\text{h}}\delta J+\phi_{\text{h}}\delta Q.
  \end{split}
\end{equation}
Clearly, Eq.~\eqref{AbsorpCond} and Eq.~\eqref{ExteDestroySim} cannot be satisfied simultaneously. The result suggests that the Coulomb and centrifugal repulsion force are just strong enough to prevent the charged test particle satisfying Eq.~\eqref{ExteDestroySim} from entering the black hole. Hence, the event horizon of an extremal Kerr-Newman Taub-NUT black hole cannot be destroyed by a charged test particle.

For an initially near-extremal Kerr-Newman Taub-NUT black hole, considering the first-order contribution of the energy
 $\delta E$, angular momentum $\delta J $ and charge $\delta Q$, the conditions to destroy the black hole are
\begin{align}
\delta E &  >\phi_{\text{h}}\delta Q+ \Omega_{\text{h}}\delta J, \label{nearextrecondition1} \\
\delta E& < \Omega_{\text{eff}}\delta J+\Phi_{\text{eff}}\delta Q-E_0. \label{nearextrecondition2}
\end{align}
Only the two conditions \eqref{nearextrecondition1} and \eqref{nearextrecondition2} are satisfied simultaneously, can the near-extremal black hole be destroyed.

\begin{figure}
  \begin{center}
\subfigure[]{\includegraphics[width=7cm]{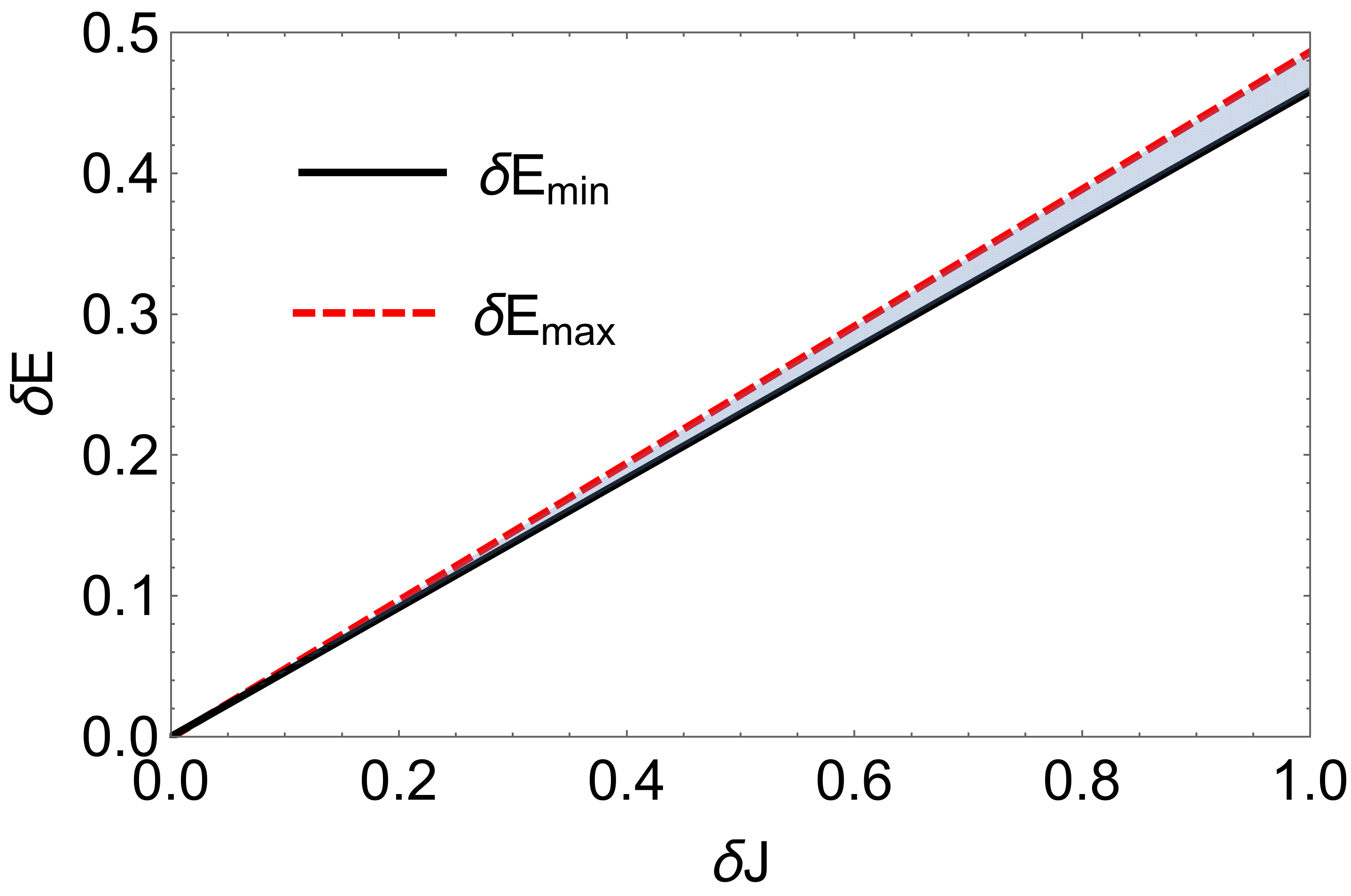}}\label{AngularmP1}
\subfigure[]{\includegraphics[width=7cm]{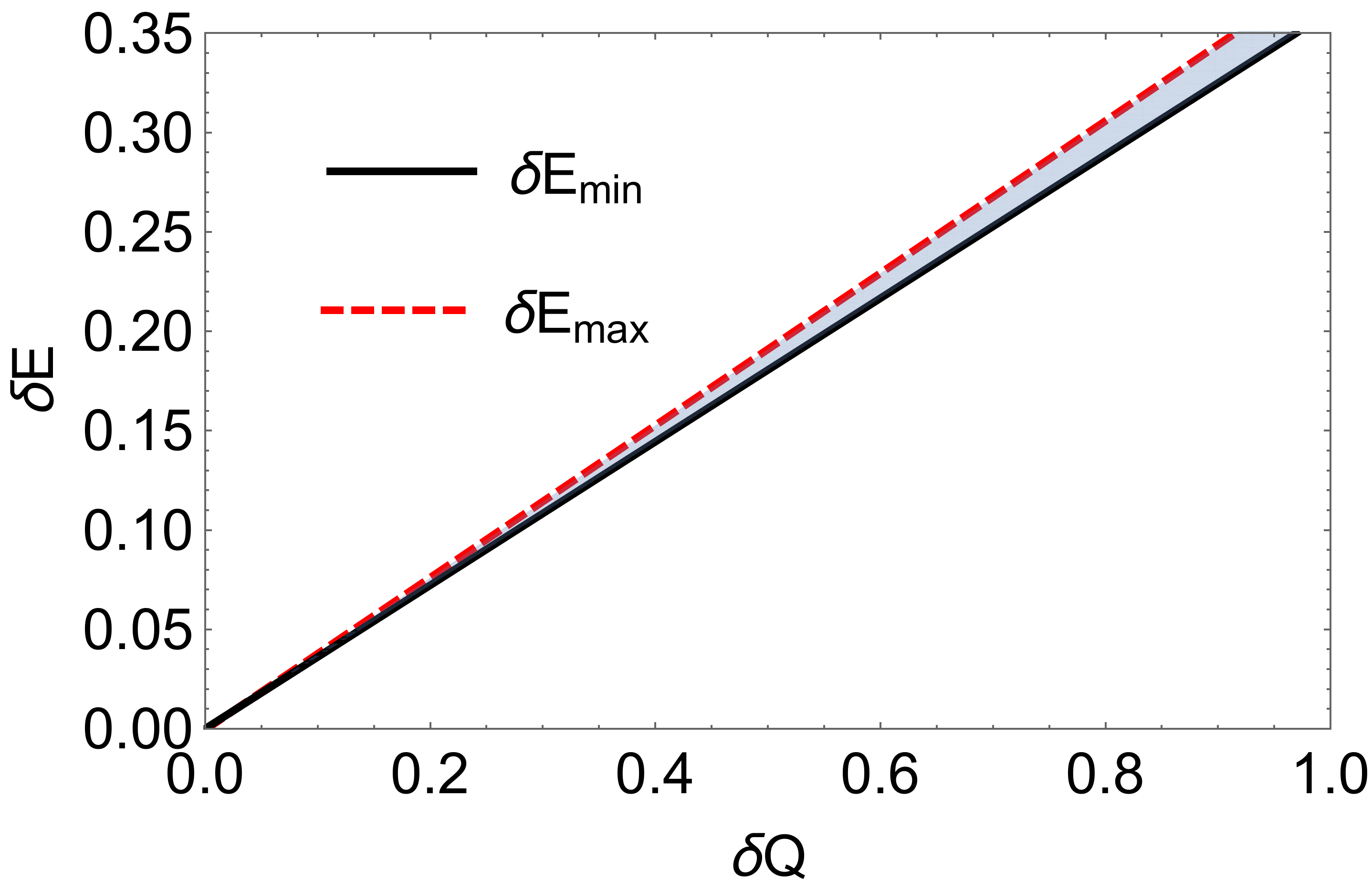}}\label{chargedParticle2}
  \caption{(color online) The energy bounds for test particles $\delta E_{\text{max}}$ (red dashed lines) and $\delta E_{\text{min}}$ (black solid lines), are plotted against the angular momentum $\delta J$ or charge $\delta Q$  of the particle for near-extremal Kerr-Newman Taub-NUT black hole with small NUT parameter. The grey regions are for $\delta E_{\text{max}}>\delta E_{\text{min}}$ that can destroy the near-extremal Kerr-Newman Taub-NUT black hole, where we have chosen the mass $M=1$ for the black hole, the NUT parameter $n=0.05$, and (a) neutral test particles injected into a near-extremal Kerr-Newman Taub-NUT black hole with angular momentum parameter $a=0.8$ and electric parameter $e=0.6$, (b) charged test particles injected along the radial direction of the near-extremal black hole, with angular momentum parameter $a=0.785$ and $e=0.62$. }\label{testparticleD}
  \end{center}
\end{figure}

In order to check whether the near-extremal Kerr-Newman Taub-NUT black hole can be destroyed by charged test particle, we define a small positive dimensionless parameter $\epsilon$ as
\begin{equation}\label{SmallPositiveParmeter}
  \frac{m^2+n^2-a^2-e^2}{m^2}=\epsilon^2.
\end{equation}
For small NUT parameter $n$, the effective angular velocity $\Omega_{\text{eff}}$ and effective electric potential $\phi_{\text{eff}}$ can be expanded as
\begin{equation}
  \begin{split}
     \Omega_{\text{eff}} & =\Omega_{\text{h}}+\frac{2 a m^2}{\left(a^2+m^2\right)^2}\epsilon+ \\
       &\frac{2 a \left(-8 a^2 m^2+a^4-m^4\right)}{3 \left(a^2+m^2\right)^4}n^2 \epsilon+\mathcal{O}(\epsilon^2), \label{EffangularV}
  \end{split}
\end{equation}
\begin{equation}
  \begin{split}
    \phi_{\text{eff}} & =\phi_{\text{h}}+\frac{m \left(m^2-a^2\right)^{3/2}}{\left(a^2+m^2\right)^2}\epsilon -\\
       & \frac{m \left(-39 a^4 m^2+23 a^2 m^4+9 a^6+7 m^6\right)}{6 \left(\sqrt{m^2-a^2} \left(a^2+m^2\right)^4\right)}n^2\epsilon+\mathcal{O}(\epsilon^2).\label{EffPotentional}
  \end{split}
\end{equation}
Evidently, for a near-extremal Kerr-Newman Taub-NUT black hole, we have
\begin{equation}
  \begin{split}
  \Omega_{\text{eff}}>\Omega_{\text{h}},  ~~~~~\phi_{\text{eff}}>\phi_{\text{h}}, ~~~~~E_{0}>0.
  \end{split}
\end{equation}
Indeed, the result indicates that when the NUT parameter is small, it is possible to destroy the near-extremal Kerr-Newman Taub-NUT black hole by charged test particle. Figure~\ref{testparticleD} shows that there exists a small range of parameters of energy, angular momentum and charge for charged test particles to destroy a near-extremal Kerr-Newman Taub-NUT black hole.

Hence, for a near-extremal black hole, there exists test particles with energy $\delta E$, angular momentum $\delta J$ and charge $\delta Q $ that can destroy the black hole.

\section{Destroying the black hole with a test charged scalar field}\label{4}

In addition to attempting to destroy the black hole using a test particle, we can also explore the possibility of destroying the black hole with classical charged test scalar field. In this section, we examine whether it is feasible to destroy the event horizon of the Kerr-Newman Taub-NUT black hole using a test complex scalar field.

\subsection{Charged scalar field in Kerr-Newman Taub-NUT spacetime}

In order to check the validity of the weak cosmic censorship conjecture with classical field, we shoot a charged test scalar field with mass $\mu_{\text{S}}$ and charge $q$ into the black hole. The action for the charged complex scalar field in the Kerr-Newman Taub-NUT spacetime can be written as
\begin{equation}\label{ActionS}
  \begin{split}
 S_{\text{S}}=\int  \sqrt{-g} d^4x \mathcal{L},
 \end{split}
 \end{equation}
where $\mathcal{L}$ is the Lagrangian density:
\begin{equation}
\begin{split}
 \mathcal{L}=-\frac{1}{2}\mathcal{D}_\mu\Psi\left(\mathcal{D}^\mu\Psi\right)^*-\frac{1}{2}\mu_{\text{S}}\Psi\Psi^*,
  \end{split}
\end{equation}
and $\mathcal{D}_\mu$ is the covariant derivative $\mathcal{ D}_\mu=\partial_\mu-iqA_\mu$. From the action, we can get the equation of motion for the complex scalar field
\begin{equation}\label{ScalarE}
  \left(\nabla_\mu-iqA_\mu\right)\left(\nabla^\mu-iqA^\mu\right)\Psi-\mu^2_{\text{S}}\Psi=0.
\end{equation}
To make the problem more tractable, it is convenient to decompose the scalar field into the form
\begin{equation}\label{ScalarFD}
  \Psi=e^{-i\omega t}R(r)S_{lm'}(\theta)e^{im'\phi}.
\end{equation}
Then we can get the angular part of the equation
\begin{multline}\label{AngularE}
 \frac{1}{\sin\theta}\frac{d}{d\theta}\left(\sin\theta\frac{dS_{lm'}}{d\theta}\right)-
  \left\{ \mu_{\text{S}}^2\left(n+a\cos\theta\right)^2+\right.\\
   \left. \frac{\left[\omega\left(2n\cos\theta-a\sin^2\theta\right)+m'\right]^2}{\sin^2\theta}
  -\lambda_{lm'}\right\}S_{lm'}(\theta)=0.
\end{multline}
and the radial equation
\begin{multline}\label{REqation}
  \frac{d}{dr}\left(\Delta \frac{dR}{dr}\right)+\Bigg\{\frac{\left[\omega\left(r^2+a^2+n^2\right)-m'a-qer\right]^2}{\Delta}  \\
 -  \mu_{\text{S}}^2r^2-\lambda_{lm'}\Bigg\}R(r)=0,
\end{multline}
where $\lambda_{lm'}$ is the separation constant and is given by $\lambda_{lm'}=l(l+1)+\mathcal{O}(a\omega)$. The solution to the angular equation is the spheroidal angular function~\cite{Seid89}. We are more concerned with the radial part since the contribution of the angular part will be reduced to unity in the fluxes by the normalization condition.

The radial part of the equation can be simplified by introducing the tortoise coordinate
\begin{equation}\label{TortoiseC}
  \frac{dr}{dr_*}=\frac{\Delta}{r^2+a^2+n^2}.
\end{equation}
Then the radial equation can be simplified as follows
\begin{multline}\label{RDequation}
  \frac{d^2R}{dr_*^2}+\frac{2r\Delta}{\left(r^2+a^2+n^2\right)}\frac{dR}{dr_*}+
  \left[\left(\omega-\frac{m'a}{r^2+a^2+n^2}  \right.\right.\\
   \left.\left.-\frac{qer}{r^2+a^2+n^2}\right)^2-
  \frac{\Delta\left(\mu_{\text{S}}^2r^2+\lambda_{lm'}\right)}{\left(r^2+a^2+n^2\right)^2}\right]R(r)=0.
\end{multline}
In the vicinity of the event horizon $r\to \rh$, where $\Delta\to 0$, the radial equation can be simplified as
\begin{equation}\label{SimEquR}
  \frac{d^2R}{dr^2_*}+\left(\omega-\frac{m'a}{\rh^2+a^2+n^2}-\frac{qer}{\rh^2+a^2+n^2}\right)^2R(r)=0.
\end{equation}
According to Eq.~\eqref{angularV} and Eq.~\eqref{Epotetional}, the above equation can be written as
\begin{equation}\label{RHEq}
  \frac{d^2R}{dr^2_*}+\left(\omega-m'\Omega_{\text{h}}-q\phi_{\text{h}}\right)^2R(r)=0.
\end{equation}
The solution to the equation is
\begin{equation}\label{HSol}
  R(r)\sim \exp\left[\pm i\left(\omega-m'\Omega_{\text{h}}-q\phi_{\text{h}}\right)r_*\right].
\end{equation}
Since there is only ingoing wave mode at the horizon, we select the minus sign:
\begin{equation}\label{HSolution}
  R(r)\sim \exp\left[- i\left(\omega-m'\Omega_{\text{h}}-q\phi_{\text{h}}\right)r_*\right].
\end{equation}
Consequently, the solution for the complex scalar field in the vicinity of the black hole event horizon is given by
\begin{equation}\label{SolutionH}
  \Psi=e^{-i\omega t}\exp\left[- i\left(\omega-m'\Omega_{\text{h}}-q\phi_{\text{h}}\right)r_*\right]S_{lm'}(\theta)e^{im'\phi}.
\end{equation}

With the solution for the charged scalar field near the event horizon, we can get the flux of the scalar field into the black hole and check the validity of the weak cosmic censorship conjecture.

We shoot a monotonic complex scalar field with mode $(l,m')$ into the black hole. After the scalar field is absorbed by the black hole, the change of parameters of the black hole can be estimated from the flux.

To get the flux of the scalar field into the black hole, we start with the energy momentum tensor of the complex scalar field. From the spacetime translation invariance for the action, we can get the energy-momentum tensor of the complex scalar field, which is~\cite{Sred07}
\begin{equation}\label{EMTensor}
\begin{split}
  T^\mu_{~\nu}&=-\frac{\partial \mathcal{L}}{\partial(\partial_\mu\Psi)}\partial_\nu\Psi-\frac{\partial \mathcal{L}}{\partial(\partial_\mu\Psi^*)}\partial_\nu\Psi^*+\delta^\mu_\nu\mathcal{L} \\
  &=\frac{1}{2}\left(\mathcal{D}^{\mu}\Psi\right)^*\partial_\nu \Psi+\frac{1}{2}\mathcal{D}^\mu\Psi\partial_\nu \Psi^* \\
  &+\delta^\mu_\nu\left(-\frac{1}{2}\mathcal{D}_\alpha\Psi\left(\mathcal{D}^{\alpha}\Psi\right)^*-\frac{1}{2}\mu_{\text{S}}^2\Psi\Psi^*\right).
\end{split}
\end{equation}
Then the energy flux through the event horizon is
\begin{equation}\label{EnergyFlux}
\begin{split}
  \frac{dE}{dt}&=\int_{\mathcal{H}}T^r_{~t}\sqrt{-g}d\theta d\phi\\
  &=\omega \left(\omega-m'\Omega_{\text{h}}-q\phi_{\text{h}}\right)\left(\rh^2+a^2+n^2\right),
\end{split}
\end{equation}
and the angular momentum flux
\begin{equation}\label{AngularMFlux}
\begin{split}
  \frac{dJ}{dt}&=-\int_{\mathcal{H}}T^r_{~\phi}\sqrt{-g}d\theta d\phi\\
  &=m' \left(\omega-m'\Omega_{\text{h}}-q\phi_{\text{h}}\right)\left(\rh^2+a^2+n^2\right).
\end{split}
\end{equation}
The action for the complex scalar field is gauge invariant, this implies that there is a conserved current---the electric current, which is~\cite{ToAl14}
\begin{equation}\label{ECurrent}
  j^\mu=\frac{iq}{2}\left(\Psi^*\mathcal{D}^\mu \Psi-(\mathcal{D}^\mu \Psi)^*\Psi\right) .
\end{equation}
From the electric current, we can get the electric flux through the event horizon, which is
\begin{equation}\label{Eflux}
\begin{split}
  \frac{dQ}{dt}&=-\int_{\mathcal{H}}j^r\sqrt{-g}d\theta d\phi\\
  &=q\left(\omega-m'\Omega_{\text{h}}-q\phi_{\text{h}}\right)\left(\rh^2+a^2+n^2\right).
\end{split}
\end{equation}
In the calculation of Eqs.~\eqref{EnergyFlux}, \eqref{AngularMFlux} and \eqref{Eflux}, we used the normalization condition of the function $S_{lm'}(\theta)$ in the integration. As indicated in the work of Bekenstein~\cite{Beke73}, during the scattering process, the ratio of the angular momentum flux to the energy flux is $m/\omega$, and the ratio of the electric flux to the energy flux is $q/\omega$.

From Eqs.~\eqref{EnergyFlux}, \eqref{AngularMFlux} and \eqref{Eflux}, the fluxes are negative for wave modes satisfying $\omega<m'\Omega_{\text{h}}+q\phi_{\text{h}}$. This indicates that energy, angular momentum and charge are extracted out from the black hole. This is the so called superradiance~\cite{BrCP15}.

Then during a small time interval $dt$, the changes of the energy, angular momentum and charge of the black hole are
\begin{subequations}\label{dEdJdQ}
  \begin{align}
    d E & =\omega\left(\omega-m'\Omega_{\text{h}}-q\phi_{\text{h}}\right)\left(\rh^2+a^2+n^2\right)dt,  \label{EnergyChange} \\
     d J &= m'\left(\omega-m'\Omega_{\text{h}}-q\phi_{\text{h}}\right)\left(\rh^2+a^2+n^2\right)dt, \label{AngularMChange}  \\
          dQ & =q\left(\omega-m'\Omega_{\text{h}}-q\phi_{\text{h}}\right)\left(\rh^2+a^2+n^2\right)dt. \label{ChargeChange}
  \end{align}
\end{subequations}

Having the changes of the mass, angular momentum and charge of the black hole during the scattering process, we can check the validity of the weak cosmic censorship conjecture for the charged scalar field scattering.

\subsection{Destroying the black hole with a { monotonic} charged scalar field}

In this subsection, we try to destroy the extremal and near-extremal Kerr-Newman Taub-NUT black holes by shooting a monotonic classical charged test scalar field with frequency $\omega$ and azimuthal harmonic index $m'$ into the black hole, and investigate the effect of the NUT parameter on the validity of the weak cosmic censorship conjecture.

Without loss of generality, let's consider a small time interval $dt$. To analyze a long-time scattering process, we divide it into a series of small time intervals $dt$ and examine each interval separately only by changing the initial parameters of the black hole.

In the scattering process, an initial extremal or near-extremal Kerr-Newman Taub-NUT black hole with mass $M$, angular momentum $J$ and charge $Q$ absorbs a complex scalar field with energy $dE$, angular momentum $dJ$, and charge $dQ$ and becomes a composite object with mass $M'$, angular momentum $J'$ and charge $Q'$. The parameters of the black hole change as
\begin{equation}\label{ParameterChangeS}
  \begin{split}
     M &\to M'=M+d E,  ~~~~~~~~~~~J \to J'=J+d J,     \\
      Q & \to Q'=Q+d Q, ~~~~~~~~~~~N \to N'=N,
  \end{split}
\end{equation}
and the metric function of the black hole changes as
\begin{equation}\label{MFuncS}
  \Delta_{\text{min}}(M,J,Q,N) \to \Delta_{\text{min}}(M+dM,J+dJ,Q+dQ,N).
\end{equation}

To check the validity of the weak cosmic censorship conjecture, we only need to check the sign of the minimal of the metric function, which is
\begin{equation}\label{MFunctionS}
\begin{split}
  \Delta'_{\text{min}}&=\Delta_{\text{min}}(M+dM,J+dJ,Q+dQ,N) \\
  &=\Delta_{\text{min}}+\frac{\partial \Delta_{\text{min}}}{\partial M}d E+\frac{\partial \Delta_{\text{min}}}{\partial J}d J+\frac{\partial \Delta_{\text{min}}}{\partial Q}d Q.
  \end{split}
\end{equation}
Substituting Eqs.~\eqref{EnergyChange}, \eqref{AngularMChange} and \eqref{ChargeChange} into Eq.~\eqref{MFunctionS}, we obtain
\begin{equation}\label{MFunctionSf}
\begin{split}
  \Delta'_{\text{min}}&=\Delta_{\text{min}}+\frac{\partial \Delta_{\text{min}}}{\partial M}d E+\frac{\partial \Delta_{\text{min}}}{\partial J}d J+\frac{\partial \Delta_{\text{min}}}{\partial Q}d Q \\
  &=-\left(m^2+n^2-a^2-e^2\right)+\left(\frac{\partial \Delta_{\text{min}}}{\partial M}\omega+\frac{\partial \Delta_{\text{min}}}{\partial J}m'  \right. \\
  &\left.+\frac{\partial \Delta_{\text{min}}}{\partial Q}q\right)\left(\omega-m'\Omega_{\text{h}}-q\phi_{\text{h}}\right)\left(\rh^2+a^2+n^2\right)dt.
  \end{split}
\end{equation}

For an extremal black hole, as Eq.~\eqref{ExteDestroyp} indicated, Eq.~\eqref{MFunctionSf} becomes
\begin{multline}\label{ExteDestroy}
   \Delta'_{\text{min}}= -\frac{\Theta}{\Upsilon}\left( m^2+n^2+a^2 \right)\left(\omega-m'\Omega_{\text{h}}-q\phi_{\text{h}}\right)^2\\
    \times\left(\rh^2+a^2+n^2\right)dt<0,
\end{multline}
where $\Upsilon $ and $\Theta$ are defined as Eq.~\eqref{DUpsi} and  Eq.~\eqref{DTheta}, respectively.
It is evident that an extremal Kerr-Newman Taub-NUT black hole cannot be destroyed by test charged scalar field.

In Sec.~\ref{3}, we introduced a small positive dimensionless parameter in Eq.~\eqref{SmallPositiveParmeter} for the particle injection to destroy the black hole. Similarly, here we also define a dimensionless parameter $\epsilon$ as
\begin{equation}\label{smallpositiveparameter}
  \frac{m^2+n^2-a^2-e^2}{m^2}=\epsilon^2.
\end{equation}

For a near-extremal Kerr-Newman Taub-NUT black hole, we have
\begin{equation}
  \begin{split}
   \Delta'_{\text{min}}
  &=-\left(m^2+n^2-a^2-e^2\right)+\left(\frac{\partial \Delta_{\text{min}}}{\partial M}\omega+\frac{\partial \Delta_{\text{min}}}{\partial J}m' \right. \\
  &\left.+\frac{\partial \Delta_{\text{min}}}{\partial Q}q\right)\left(\omega-m'\Omega_{\text{h}}-q\phi_{\text{h}}\right)\left(\rh^2+a^2+n^2\right)dt  \\
  &=-\left(m^2+n^2-a^2-e^2\right)+\frac{\partial \Delta_{\text{min}}}{\partial M}\left(\omega-m'\Omega_{\text{eff}} \right.\\
  &\left.-q\phi_{\text{eff}}\right) \left(\omega-m'\Omega_{\text{h}}-q\phi_{\text{h}}\right)\left(\rh^2+a^2+n^2\right)dt.
  \end{split}
\end{equation}
The above equation can be regarded as a quadratic equation for the frequency $\omega$ of the complex scalar field.
If the complex scalar field shotting into black hole satisfies
\begin{equation}\label{FieldMod}
  \omega=\frac{m'\left(\Omega_{\text{eff}}+\Omega_{\text{h}}\right)+q\left(\phi_{\text{eff}}+\phi_{\text{h}}\right)}{2},
\end{equation}
the minimal of the metric function is the largest. If these modes cannot destroy the event horizon of the black hole, all the modes cannot destroy the black hole, either.

After these modes of the charged scalar field are absorbed by the near-extremal black hole, the minimal of the metric function is
\begin{equation}
  \begin{split}
   \Delta'_{\text{min}}
  &=-\left(m^2+n^2-a^2-e^2\right)+\frac{\partial \Delta_{\text{min}}}{\partial M}\left(\omega-m'\Omega_{\text{eff}}\right. \\
  &\left.-q\phi_{\text{eff}}\right)\left(\omega-m'\Omega_{\text{h}}-q\phi_{\text{h}}\right)\left(\rh^2+a^2+n^2\right)dt\\
   &=-\left(m^2+n^2-a^2-e^2\right)-\frac{1}{4}\frac{\partial \Delta_{\text{min}}}{\partial M}\left[m'\left(\Omega_{\text{eff}} \right. \right. \\
   &\left.\left.-\Omega_{\text{h}}\right)+
   q\left(\phi_{\text{eff}}-\phi_{\text{h}}\right)\right]^2\left(\rh^2+a^2+n^2\right)dt.\label{ScalarMinimal}
  \end{split}
\end{equation}
As indicated by Eq.~\eqref{EffangularV} and Eq.~\eqref{EffPotentional}, for small NUT parameter, we have
\begin{equation}
  \Omega_{\text{eff}}-\Omega_{\text{h}} =\frac{2 a m^2}{\left(a^2+m^2\right)^2}\epsilon+\frac{2 a \left(-8 a^2 m^2+a^4-m^4\right)}{3 \left(a^2+m^2\right)^4}n^2 \epsilon, \label{Omegaeff2}
\end{equation}
\begin{multline}
  \phi_{\text{eff}} -\phi_{\text{h}} =\frac{m \left(m^2-a^2\right)^{3/2}}{\left(a^2+m^2\right)^2}\epsilon \\
  -\frac{m \left(-39 a^4 m^2+23 a^2 m^4+9 a^6+7 m^6\right)}{6 \left(\sqrt{m^2-a^2} \left(a^2+m^2\right)^4\right)}n^2\epsilon. \label{Phieff2}
\end{multline}
Plugging Eqs.~\eqref{Omegaeff2} and \eqref{Phieff2} into Eq.~\eqref{ScalarMinimal}, we can obtain
\begin{equation}\label{ScalarMinimalSimplified}
  \begin{split}
      \Delta'_{\text{min}}&  =-m^2\epsilon^2-\frac{1}{4}\frac{\partial \Delta_{\text{min}}}{\partial M}\left\{ m'\left[\frac{2 a m^2}{\left(a^2+m^2\right)^2} \right.\right.+ \\
       & \left.  \frac{2 a \left(-8 a^2 m^2+a^4-m^4\right)}{3 \left(a^2+m^2\right)^4}n^2\right]
        \left. + q\left[\frac{m \left(m^2-a^2\right)^{3/2}}{\left(a^2+m^2\right)^2} \right. \right.\\
       & \left.\left.-\frac{m \left(-39 a^4 m^2+23 a^2 m^4+9 a^6+7 m^6\right)}{6 \left(\sqrt{m^2-a^2} \left(a^2+m^2\right)^4\right)}n^2\right]\right\}^2  \\
       &\times\epsilon^2\left(\rh^2+a^2+n^2\right)dt.
  \end{split}
\end{equation}
%
%
Since $dt$ is of order $\epsilon$, then for small NUT parameter $n$, we have
\begin{equation}\label{ScalarDeltaMin}
\Delta'_{\text{min}}
   =-m^2\epsilon^2+\mathcal{O}(\epsilon^3)<0.
\end{equation}
It is clear that a near-extremal Kerr-Newman Taub-NUT black hole cannot be destroyed by test complex scalar field.

Consequently, our investigation suggests that both extremal and near-extremal Kerr-Newman Taub-NUT black hole cannot be destroyed by charged test scalar field.

\subsection{Destroying the black hole with a non-monotonic charged scalar field}

Having considered the gedanken experiment to destroy a black hole by a monotonic charged scalar field, now we consider to destroy the black hole by shooting a more complex charged scalar field.

We consider to shoot a charged scalar field which is a superposition state of approximate solutions corresponding to two different frequencies $\omega$. The solution for the scalar field in the vicinity of the black hole event horizon is
\begin{equation}\label{psinonmonotonic}
  \begin{split}
      \Psi&=Ae^{-i\omega_1 t}\exp\left[- i\left(\omega_1-m'\Omega_{\text{h}}-q\phi_{\text{h}}\right)r_*\right]S_{lm'}(\theta)e^{im'\phi} \\
       & +Be^{-i\omega_2 t}\exp\left[- i\left(\omega_2-m'\Omega_{\text{h}}-q\phi_{\text{h}}\right)r_*\right]S_{lm'}(\theta)e^{im'\phi}.
  \end{split}
\end{equation}

Following the same procedure as the previous subsection, we can get the changes of mass, angular momentum and charge of the black hole during the small time interval $d t$, which are

\begin{subequations}
 \begin{equation}
  \begin{split}
     dE & =\left\{\omega_1\left(\omega_1-m'\Omega_{\text{h}}-q\phi_{\text{h}}\right)\mid A\mid^2+\omega_2\left(\omega_2-m'\Omega_{\text{h}}\right.\right. \\
       &  \left. -q\phi_{\text{h}}\right)\mid B \mid^2 +\left[ \omega_1\left(\omega_2-m'\Omega_{\text{h}}-q\phi_{\text{h}}\right)+
       \omega_2\left(\omega_1 \right.\right.  \\
       &\left.\left.\left.-m'\Omega_{\text{h}}-q\phi_{\text{h}} \right)\right]\text{Re}\left[ AB^*\exp[-i(\omega_1-\omega_2)(t+r_*)]\right] \right\} \\
       &\times\left(\rh^2+a^2+n^2\right)dt,
  \end{split}
\end{equation}
\begin{equation}
  \begin{split}
     dJ & =m'\left\{\left(\omega_1-m'\Omega_{\text{h}}-q\phi_{\text{h}}\right)\mid A\mid^2+\left(\omega_2-m'\Omega_{\text{h}}\right.\right. \\
       &\left.  \left. -q\phi_{\text{h}}\right)\mid B \mid^2 +\left( \omega_1+\omega_2-2m'\Omega_{\text{h}}-2q\phi_{\text{h}}\right)\right.\\
     & \left. \times\text{Re}\left[ AB^*\exp[-i(\omega_1-\omega_2)(t+r_*)]\right] \right\}\left(\rh^2+a^2+n^2\right)dt,
  \end{split}
\end{equation}
\begin{equation}
  \begin{split}
     dQ & =q\left\{\left(\omega_1-m'\Omega_{\text{h}}-q\phi_{\text{h}}\right)\mid A\mid^2+\left(\omega_2-m'\Omega_{\text{h}}\right.\right. \\
       &\left.  \left. -q\phi_{\text{h}}\right)\mid B \mid^2 +\left( \omega_1+\omega_2-2m'\Omega_{\text{h}}-2q\phi_{\text{h}}\right)\right.\\
     & \left. \times\text{Re}\left[ AB^*\exp[-i(\omega_1-\omega_2)(t+r_*)]\right] \right\}\left(\rh^2+a^2+n^2\right)dt.
  \end{split}
\end{equation}
\end{subequations}

For an extremal Kerr-Newman Taub-NUT black hole, we only need to check the sign of the minimal of the metric function $\Delta'_{\text{min}}$. We find
\begin{equation}
\begin{split}
   \Delta'_{\text{min}}&= -\frac{\left( m^2+n^2+a^2 \right)\Theta}{\Upsilon}\left(d E-\Omega_{\text{h}}d J-\phi_{\text{h}}d Q    \right) \\
   &\leq-\frac{\left( m^2+n^2+a^2 \right)\Theta}{\Upsilon}
   \left[
   \mid\omega_1-m'\Omega_{\text{h}}-q\phi_{\text{h}}\mid\mid A\mid \right. \\
   &\left.- \mid\omega_2-m'\Omega_{\text{h}}-q\phi_{\text{h}}\mid\mid B\mid
   \right]^2\left(\rh^2+a^2+n^2\right)dt<0.
\end{split}
\end{equation}
The result shows that the charged scalar field which is the superposition state of approximate
solutions corresponding to two different $\omega$ cannot destroy an extremal Kerr-Newman Taub-NUT black hole.

In fact, applying the same procedure for the charged scalar field which is the superposition state of approximate
solutions corresponding to any different frequencies $\omega$,
\begin{equation}
  \Psi=\sum\limits_{i} A_i e^{-i\omega_i t}e^{ - i\left(\omega_i-m'\Omega_{\text{h}}-q\phi_{\text{h}}\right)r_*} S_{lm'}(\theta)e^{im'\phi}
\end{equation}
we can show that the minimal of the metric fuction
\begin{equation}
\begin{split}
   \Delta'_{\text{min}}&= -\frac{\left( m^2+n^2+a^2 \right)\Theta}{\Upsilon}\left(d E-\Omega_{\text{h}}d J-\phi_{\text{h}}d Q    \right) \\
   &\leq-\frac{\left( m^2+n^2+a^2 \right)\Theta}{\Upsilon}
   \Bigg[
   \sum\limits_{i}\epsilon_i\mid\omega_i-m'\Omega_{\text{h}}  \\
   &
   -q\phi_{\text{h}}\mid\mid A_i\mid
   \Bigg]^2\left(\rh^2+a^2+n^2\right)dt<0,
\end{split}
\end{equation}
where $\epsilon_i=\pm 1$.
The result suggests that an extremal Kerr-Newman Taub-NUT black hole cannot be destroyed by a non-monotonic charged scalar field.


\section{Discussion and Conclusions}\label{7}

Spacetime singularities are windows onto physics beyond general relativity, and weak cosmic censorship conjecture has become one of the foundations of black hole physics. In this paper, we investigated the weak cosmic censorship conjecture for the Kerr-Newman Taub-NUT black hole by throwing a charged test particle and a charged scalar field. However, to investigate the weak cosmic censorship conjecture using gadenken experiments, the key problem is how do the parameters of the black hole change. This problem is closely related to the first law of black hole thermodynamics. Due to the presence of the Misner strings, the Kerr-Newman Taub-NUT spacetime is not asyptotic flat. This leads to the non-uniqueness for the definition of thermodynamic quantities such as mass, angular momentum, and charge.  Following the viewpoint of Hennigar et al., we constructed the first law of black hole thermodynamics for the Kerr-Newman Taub-NUT black hole. Different from other black hole solutions, the electric charge in the first law is defined as the Komar integral at the event horizon instead of infinity.

Having the first law, we investigated the weak cosmic censorship conjecture by a charged test particle and a complex scalar field. We found that particles causing the destruction of the extremal Kerr-Newman Taub-NUT black hole cannot be absorbed by the black hole. However, there exists particles with energy, charge and/or angular momentum to destroy the near-extremal Kerr-Newman Taub-NUT black hole. For charged scalar field scattering, we found that both the extremal and near-extremal Kerr-Newman Taub-NUT black holes cannot be destroyed by { a monotonic charged scalar field, and an extremal Kerr-Newman Taub-NUT black hole cannot be destroyed by a non-monotonic complex scalar field. }

{In our investigation, we showed that there is a small range of parameters for the energy, angular momentum and charge of the test particle that can destroyed the event horizon of a near-extremal Kerr-Newman Taub-NUT black hole. It seems that the weak cosmic censorship conjecture might not valid for this black hole, however, when blackreaction and self-force are taken into account, the event horizon of a near-extremal Kerr-Newman Taub-NUT black hole might still cannot be destroyed as the case of a near-extremal Kerr-Newman black hole~\cite{SoWa17}.
}
\acknowledgments

 This work was supported in part by the National Natural Science Foundation of China (Grants No. 12247178, No. 11875151, No. 12075103, No. 12047501, No. 12105126, No. 12205129 and No. 12247101), the China Postdoctoral Science Foundation (Grant No. 2023M731468 and 2021M701529), the 111 Project (Grant No. B20063), the Major Science and Technology Projects of Gansu Province, and Lanzhou City's scientific research funding subsidy to Lanzhou University.\\
~\\
\textit{Note added.}—At the time we finished our work, we were aware of a manuscript~\cite{Wu23} appeared in arXiv three days before our manuscript which independently derives the same result for the first law of black hole thermodynamics for Kerr-Newman Taub-NUT black hole. We reported our main results on Apr. 23 (2023) in the Annal Meeting of Division of Gravity and Relativistic Astrophysics of China, and we also reported our results on May 17 (2023) in the Asia-Pacific School and Workshop on Gravitation and Cosmology 2023, and the slide can be found on the APSW-GC website: https://indico.ictp-ap.org/event/77/.

%
\end{document}